\documentclass[12pt]{article}
\usepackage[dvips]{graphicx}
\usepackage{dcolumn}
\usepackage{bm}

\newcommand{\be}{\begin{eqnarray} }
\newcommand{\ee}{\end{eqnarray} }
\newcommand{\beq}{\begin{equation} }
\newcommand{\eeq}{\end{equation} }

\newcommand{\palka}{{\Bigl |}}
\begin{document} 
\begin{center}{\large Transversity and its accompanying T-odd distribution from 
Drell-Yan processes with pion-proton collisions\vskip 1cm
{ A. Sissakian, O. Shevchenko, A. Nagaytsev, O. Denisov,O. Ivanov\\
{\it  Joint Institute for Nuclear Research, 141980 Dubna, Russia}
}}
\end{center}
\begin{abstract}
It is studied  the possibility of direct extraction of the transversity and its
accompanying T-odd parton distribution function (PDF)  from   Drell-Yan (DY) processes 
with unpolarized pion beam and with both unpolarized
and transversely polarized proton targets. At present, such a  measurement can be performed on 
the COMPASS experiment at CERN.
The preliminary estimations performed for COMPASS
kinematic region demonstrate that it is quite real to extract both  transversity and its accompanying T-odd PDF
in the COMPASS conditions.
\end{abstract}
The advantage of DY process for extraction of PDFs, 
is that there is no need in any fragmentation functions.
It is well known that  the double transversely polarized DY process
$H_1^{\uparrow}H_2^{\uparrow} \to l^{+}l^{-}X$  allows  to directly extract the transversity distributions 
(see Ref. \cite{ba} for review). In particular, the double polarized DY process with antiproton beam
is planned to be studied at PAX \cite{pax}. However, this is a rather difficult task to produce antiproton
beam with the sufficiently high degree of polarization. So, it is certainly desirable to have
an alternative (complementary) possibility allowing to extract the transversity PDF from unpolarized and single-polarized
DY processes.  This could be a matter of especial interest for the COMPASS experiment \cite{comp} where
the possibility to study DY processes with unpolarized pion beam
and with both unpolarized and transversely polarized proton targets 
$\pi^-p\rightarrow\mu^+\mu^-X$,  $\pi^-p^\uparrow\rightarrow\mu^+\mu^-X$ 
is under discussion now.

The original expressions for unpolarized and single-polarized DY cross-sections \cite{bo1} are very
inconvenient in application since all $k_T$-dependent PDFs enter there in the complex convolution.
To avoid this problem in Ref. \cite{approach} the $q_T$ integration approach \cite{mul1,mul2,mul3} was applied.
As a result, the procedure proposed in Ref. \cite{approach} allows to extract the transversity $h_1$ and the
first moment 
\be
\label{hperpmom}
h_{1q}^{\perp(1)}(x)\equiv\int d^2{\bf k}_T\left(\frac{{\bf k}_T^2}{2M_\pi^2}\right)h_{1q}^\perp(x_\pi,{\bf k}_T^2)
\ee
of T-odd distribution $h_1^\perp$ directly, without any model assumptions about $k_T$-dependence 
of $h_1^\perp(x,k_T^2)$.

The general procedure proposed in Ref. \cite{approach} applied to unpolarized DY process $\pi^-p\rightarrow\mu^+\mu^-X$
gives\footnote{Eq. (\ref{e1}) is obtained within the  quark parton model. It is of importance that 
the large values of $\nu$ cannot be explained by leading and next-to-leading order perturbative QCD
corrections as well as by the high twists effects (see \cite{bo1} and references therein).} 
\be
\label{e1}
\hat k\palka_{\pi^-p\rightarrow \mu^+\mu^-X} =8\frac{\sum_qe_q^2[\bar h_{1q}^{\perp(1)}(x_{\pi}){\Bigl |}_{\pi^-} h_{1q}^{\perp(1)}(x_p){\Bigl |}_p+(x_\pi\leftrightarrow x_p)]}
{\sum_qe_q^2[\bar f_{1q}(x_\pi){\Bigl |}_{\pi^-}f_{1q}(x_p){\Bigl |}_{p}+(x_\pi\leftrightarrow x_p)]},
\ee
where $\hat k$ is the coefficient at $\cos 2\phi$ dependent part of the properly integrated over $q_{T}$ 
ratio of unpolarized cross-sections:
\be
\label{r1}
& & \hat R=\frac{\int d^2 {\bf q}_T [{\bf |}{\bf q}_T{\bf |}^2/{M_{\pi}M_p}][d\sigma^{(0)}/d\Omega]}{\int d^2{\bf q}_T\sigma^{(0)}},\\
\label{r2}
 &  & \hat R=\frac{3}{16\pi}(\gamma(1+\cos^2\theta)+\hat k\cos 2\phi\sin^2\theta).
\ee
At the same time in the case of single polarized DY process  $\pi^-p^\uparrow\rightarrow\mu^+\mu^-X$, operating just as in Ref. \cite{approach},
one gets
\be
\hat A_{h}  & = &
-\frac{1}{2} \frac{\sum_q e_q^2[
\bar h_{1q}^{\perp(1)}(x_\pi) h_{1q}(x_p)+(x_\pi\leftrightarrow x_p)]}
{\sum_q e_q^2 [\bar f_{1q}(x_\pi) f_{1q}(x_p)+(x_\pi\leftrightarrow x_p)]},
\ee
where the single spin asymmetry (SSA) $\hat A_h$ is defined as\footnote{Notice that SSA $\hat A_h$
is analogous to  asymmetry 
$A_{UT}^{\sin(\phi-\phi_S)\frac{q_T}{M_N}}$ (weighted with $\sin(\phi-\phi_S)$ and the same weight
$q_T/M_N$) applied in Ref. \cite{efremov} with respect to Sivers function extraction from the single-polarized DY processes.}
\be
\label{ee2}
\hat A_{h}=\frac{\int d\Omega d\phi_{S_2}\int d^2{\bf q}_T(|{\bf q}_T|/M_\pi)\sin(\phi+\phi_{S_2})[d\sigma({\bf S}_{2T})-d\sigma(-{\bf S}_{2T})]}{\int d\Omega d \phi_{S_2}\int d^2{\bf q}_T[d\sigma({\bf S}_{2T})+d\sigma(-{\bf S}_{2T})]}.
\ee
In Eqs. (\ref{e1}-\ref{ee2}) the quantity $h_{1q}^{\perp(1)}(x_\pi)$ is defined by Eq. (\ref{hperpmom}).
All other notations are the same as in Ref. \cite{approach} (see Ref. \cite{ba} for details on kinematics in the Collins-Soper frame we deal with). 

Neglecting strange quark PDF contributions, squared sea contributions of $u$-quark PDF 
$h_{1u}^{\perp(1)}(x_\pi)\palka_{\pi^-}\bar h_{1u}^{\perp(1)}(x_p)\palka_{p}$, $f_{1u}(x_\pi)\palka_{\pi^-}\bar f_{1u}(x_p)\palka_{p}$,
and cross terms containing the products of sea and valence $d$-quark PDFs (additionally suppressed by the charge factor $1/4$),
one arrives at the simplified equations
\be
\label{comk}
\hat k(x_\pi,x_p)\palka_{\pi^-p} \simeq8\frac{\bar h_{1u}^{\perp(1)}(x_{\pi}){\Bigl |}_{\pi^-} h_{1u}^{\perp(1)}(x_p){\Bigl |}_p}
{\bar f_{1u}(x_\pi){\Bigl |}_{\pi^-}f_{1u}(x_p){\Bigl |}_{p}},\\
\label{coma}
\hat A_{h}(x_\pi,x_p)\palka_{\pi^-p^\uparrow}  \simeq
-\frac{1}{2} \frac{
\bar h_{1u}^{\perp(1)}(x_\pi){\Bigl |}_{\pi^-} h_{1u}(x_p){\Bigl |}_{p}}
{\bar f_{1u}(x_\pi)\palka_{\pi^-} f_{1u}(x_p)\palka_p}.
\ee
Notice that while two equations corresponding to 
unpolarized and single-polarized antiproton-proton DY processes completely determine the transversity and its accompanying 
T-odd PDF in proton \cite{approach}, two Eqs. (\ref{comk}) and (\ref{coma}) contain three unknown quantities $\bar h_{1u}^{\perp(1)}(x_\pi)\palka_{\pi^-}$,
$h_{1u}^{\perp(1)}(x_p)\palka_p$ and $h_{1u}(x_p)$.
Nevertheless, from Eqs. (\ref{comk}) and (\ref{coma}) it immediately follows that 
\be
\frac{h_{1u}^{\perp(1)}(x_p)\palka_p}{h_{1u}(x_p)\palka_p}=-\frac{1}{16}
\frac{\hat k(x_\pi,x_p)\palka_{\pi^-p}}{\hat A_{h}(x_\pi,x_p)\palka_{\pi^-p^\uparrow}}
\ee
Thus, using only unpolarized pion beam colliding with unpolarized and transversely polarized protons, it is possible to 
extract the ratio ${h_{1u}^{\perp(1)}(x_p)\palka_p}\,/\,{h_{1u}(x_p)\palka_p}$. However, it is certainly 
desirable to extract ${h_{1u}^{\perp(1)}(x_p)\palka_p}$ and ${h_{1u}(x_p)\palka_p}$ in separation.

The simplest way to solve this problem is to use in Eqs. (\ref{comk}), (\ref{coma}) the quantity $h_{1u}^{\perp(1)}\palka_p$ 
extracted  from $\hat k$ measured in unpolarized DY process, $\bar pp\rightarrow l^+l^-X$ (in the way proposed in Ref. \cite{approach}).
However, if one wishes to extract all quantities within the experiment with the pion beam (COMPASS here) the additional assumptions connecting pion
and proton PDFs are necessary. Taking into account the probability interpretations of $h_{1q}^\perp$ and $f_{1q}$ PDFs, it is natural
to write the relation 
\be
\label{assumption}
\frac{\bar h_{1u}^{\perp(1)}(x)\palka_{\pi^-}}{h_{1u}^{\perp(1)}(x)\palka_{p}}=C_u\frac{\bar f_{1u}(x)\palka_{\pi^-}}{f_{1u}(x)\palka_p}.
\ee
Notice that the assumption given by Eq. (\ref{assumption}) is in accordance (but is much less strong restriction)
with the Boer's model (see Eq. (50) in Ref. \cite{bo1}), where $C_u=M_pc^u_\pi/M_\pi c^u_p$.

As we will see below, one should put $C_u$ to be about unity
\be
\label{cu}
C_u\simeq1
\ee
to reconcile the results on $h_{1u}^{\perp(1)}$ in proton obtained from the simulated $\hat k\palka_{\pi^-p}$ 
with the respective results \cite{approach} obtained from the  simulated  $\hat k\palka_{\bar pp}$ as well as with
the upper bound \cite{approach} on this quantity. 

Thus,   Eqs. (\ref{comk}) and (\ref{coma}) are rewritten as (c.f. Eqs. (19), (20) in Ref. \cite{approach})
\be
\label{kassumption}
\hat k(x_\pi,x_p)\palka_{\pi^-p} \simeq8\frac{ h_{1u}^{\perp(1)}(x_{\pi}){\Bigl |}_{p} h_{1u}^{\perp(1)}(x_p){\Bigl |}_p}
{f_{1u}(x_\pi){\Bigl |}_{p}f_{1u}(x_p){\Bigl |}_{p}}\\
\label{aassumption}
\hat A_{h}(x_\pi,x_p)\palka_{\pi^-p^\uparrow}  \simeq
-\frac{1}{2} \frac{
 h_{1u}^{\perp(1)}(x_\pi){\Bigl |}_{p} h_{1u}(x_p){\Bigl |}_{p}}
{ f_{1u}(x_\pi)\palka_{p} f_{1u}(x_p)\palka_p}.
\ee

Looking at Eqs. (\ref{kassumption}), (\ref{aassumption}), one can see that now the number of equations is equal to the number of variables
to be found.
Measuring the quantity $\hat k$ in unpolarized DY (Eqs. (\ref{r1}), (\ref{r2}))  and using Eq. (\ref{kassumption})
one can obtain the quantity $h_{1u}^{\perp(1)}\palka_p$. Then, measuring SSA,  Eq. (\ref{ee2}),
and using in Eq. (\ref{aassumption}) the obtained from unpolarized DY quantity $h_{1u}^{\perp(1)}\palka_p$, one can eventually extract the transversity
distribution $h_{1u}\palka_p$.

To deal with Eqs. (\ref{kassumption}) and (\ref{aassumption}) in practice, one should consider them  
at the points\footnote{The different points $x_F=0$ can
be reached changing $Q^2$ value at fixed $s\equiv Q^2/x_1x_2\equiv Q^2/\tau$} $x_\pi=x_p\equiv x$
(i.e., $x_F\equiv x_\pi-x_p=0$), so that 
\be
\label{perp}
h^{\perp(1)}_{1u}(x)=f_{1u}(x)\sqrt{\frac{\hat k(x,x)\palka_{\pi^-p}}{8}},
\ee  
and 
\be
\label{trans}
h_{1u}(x)=-4\sqrt{2}\frac{\hat A_h(x,x)\palka_{\pi^-p^\uparrow}}{\sqrt{\hat k(x,x)\palka_{\pi^-p}}}f_{1u}(x),
\ee
where now all PDFs {\it refer to proton}.

\begin{figure}[t]
\begin{center}
\includegraphics[height=5cm,width=8cm]{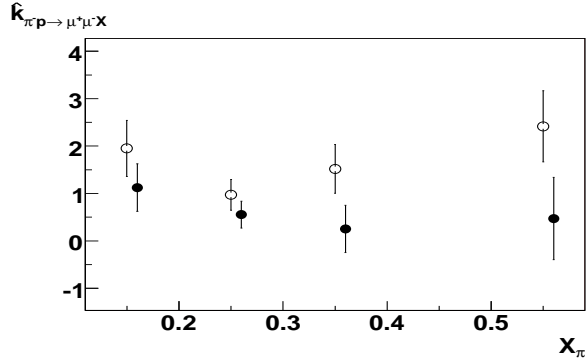}
\end{center}
\caption{\it $\hat k$  versus $x_\pi$ for $x_F\simeq0$. Data is obtained with MC simulations for
60 GeV (closed circles) and 100 GeV (open circles) pion beams. }
\label{k-hat}       
\end{figure}

To estimate the possibility of 
$h_{1u}^{\perp(1)}$  measurement, 
the special simulation of unpolarized DY
events with the  COMPASS kinematics  are performed. The pion-proton collisions are
simulated with the PYTHIA generator \cite{pythia}. Two samples are prepared
corresponding to 60 GeV and 100 GeV pion beams.
 Each sample contains about 100~K
 pure Drell-Yan events. The events are weighted (see Ref. \cite{approach} for detail) with  the ratio of DY cross-sections
given by (see Refs. \cite{bo1,conway})
\be
\label{r3}
&& R\equiv \frac{d\sigma^{(0)}/d\Omega}{\sigma^{(0)}},\\
\label{r4}
&&  R=\frac{3}{16\pi}(1+\cos^2\theta+ (\nu/2)\cos 2\phi\sin^2\theta) 
\quad(\nu \equiv 2\kappa ),
\ee
where $\nu$ dependencies of $q_T$ and $x_\pi$ are taken from Refs. \cite{conway,NA10}. 
\begin{figure}[h!]
\begin{center}
\includegraphics[height=5cm,width=8cm]{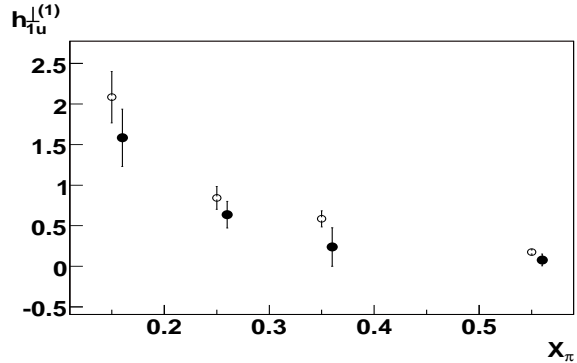}
\end{center}
\caption{\it $h_{1u}^{\perp(1)}$  versus $x_\pi$ for $x_F\simeq0$. Data is obtained with MC simulations for
60 GeV (closed circles) and 100 GeV (open circles) pion beams.}
\label{hp-hat}       
\end{figure}

The angular distributions of $\hat R$ (Eqs. (\ref{r1}) and (\ref{r2})) for both samples  
are studied 
just as it was done  in Ref. \cite{conway} with respect to $R$ (Eqs. (\ref{r3}), (\ref{r4})). 
The results are shown in Fig. \ref{k-hat}. 
The value of $\hat k$ at averaged Q$^2$ for both energies are found to be $0.7 \pm 0.1$ for
60 GeV and  $0.9 \pm 0.1$ for 100 GeV pion beams. 

The quantity $h_{1u}^{\perp (1)}$ is reconstructed 
from the obtained values of $\hat k$
using Eq. (\ref{perp}) with $x_F=0 \pm 0.04$. The results are shown in Fig. \ref{hp-hat}.
Let us recall that to obtain $h_{1u}^{\perp(1)}$ from $\hat k\palka_{\pi^- p}$, we chosen
$C_u\simeq 1$. Notice that namely this choice of $C_u$ is consistent with the results on $h_{1u}^{\perp(1)}$
obtained in Ref. \cite{approach} from simulated $\hat k\palka_{\bar pp}$ (compare Fig. \ref{hp-hat} with Fig. 4 of Ref. \cite{approach}), 
and also with the upper bound
on $h_{1u}^{\perp(1)}$ estimated in that paper. Otherwise, if $C_u$ essentially differs from unity,
one should multiply the results on $h_{1u}^{\perp(1)}$ by the factor $1/\sqrt{C_u}$,
that would lead to disagreement of the results on $h_{1u}^{\perp(1)}$ obtained from the simulated quantities $\hat k\palka_{\pi^-p}$ and
$\hat k\palka_{\bar pp}$.

Certainly, all conclusions made on the basis of simulations are very 
preliminary. The reliable conclusion about $C_u$ can be made only from the future 
measurements of $\hat k$ for both DY processes with $\bar p$ and $\pi^-$ participation.

\begin{figure}[t]
\begin{center}
\includegraphics[height=5cm,width=8cm]{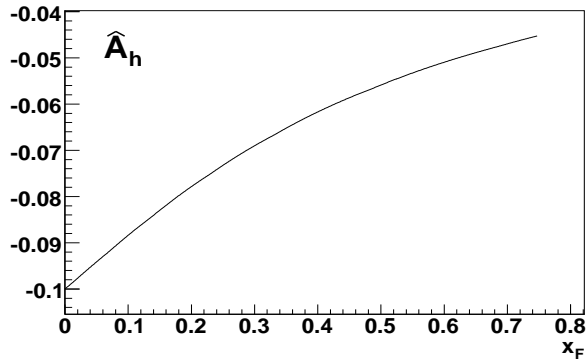}
\end{center}
\caption{\it SSA given by Eq. (\ref{aassumption}) versus $x_F$ for 100 GeV pion beam ($Q^2_{average}=6.2\,GeV^2$).}
\label{ssa-collider}       
\end{figure}

\begin{figure}[h!]
\begin{center}
\includegraphics[height=5cm,width=8cm]{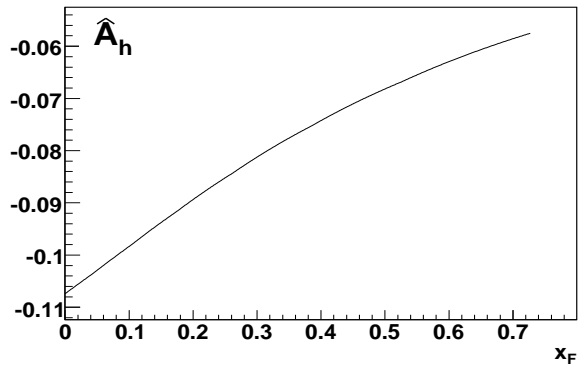}
\end{center}
\caption{\it SSA given by Eq. (\ref{aassumption}) versus $x_F$ for 60 GeV pion beam ($Q^2_{average}=5.5\,GeV^2$).}
\label{ssa-fixed}       
\end{figure}

Using the obtained magnitudes of $h_{1u}^{\perp (1)}$ we estimate the expected SSA given by Eq. (\ref{aassumption}).
The results are shown in Figs. \ref{ssa-collider} and  \ref{ssa-fixed}. 
For estimation of $h_{1u}$ entering SSA together with $h_{1u}^{\perp (1)}$ (see Eq. (\ref{aassumption})) we follow the procedure of Ref. \cite{anselmino1} 
and use (rather crude) ``evolution model''
\cite{ba, anselmino1} , where there is no any estimations of uncertainties. That is why in (purely 
qualitative) figures \ref{ssa-collider}  and \ref{ssa-fixed} we present the solid curves instead of points with error bars.
To obtain these curves we reproduce  $x$-dependence of $h_{1u}^{\perp(1)}$  in the considered region, 
using the Boer's model (Eq. (50) in Ref. \cite{bo1}),  
properly numerically corrected in accordance with the simulation results.

It should be noticed that  the estimations 
of $\hat k$ and $\hat A_h$ magnitudes obtained it this paper
are very preliminary and show just the order of values of these quantities.
For more precise estimations one needs the Monte-Carlo generator   
more suitable for DY processes studies (see, for example. Ref. \cite{bianconi} ) than PYTHIA generator which we used 
(with the proper weighting of events) here.

In summary, it is shown that  the proposed in Ref. \cite{approach} procedure can be applied  to DY processes: $\pi^-p \to \ \mu^+ \mu^- X$ and
$\pi^-p^{\uparrow} \to \ \mu^+ \mu^-  X$, which could be studied
in the COMPASS experiment at CERN. The preliminary estimations for COMPASS kinematical region show the possibility  to measure both $\hat k$ and SSA $\hat A_h$ and
then to extract  the quantities $h_1^{\perp (1)}$ and  $h_1$ we are interesting in.

  The authors are grateful to M.~Anselmino, F.~Balestra, R.~Bertini, M.P. Bussa,  A.~Efremov, L.~Ferrero, V.~Frolov, T.~Iwata,  V.~Krivokhizhin,
 A.~Kulikov, P.~Lenisa, A. Maggiora, A.~Olshevsky, 
D. Panzieri, G. Piragino, G.~Pontecorvo, I.~Savin, M.~Tabidze, O.~Teryaev, W.~Vogelsang and also to all members of Compass-Torino group
 for fruitful discussions. 
 The work  of O.S. and O.I. was supported by the Russian Foundation
 for Basic Research (project no. 05-02-17748).


\begin{thebibliography}{99}
        
\bibitem{ba} V. Barone, A. Drago, and P.G. Ratcliffe,  Phys. Rep. {\bf 359}, 1 (2002). 
\bibitem{pax} V. Barone et. al (PAX collaboration), hep-ex/0505054 
\bibitem{comp} G.K.Mallot, Nucl. Inst. Meth. A518 (2004) 121.

\bibitem{bo1} D. Boer, Phys. Rev. {\bf D}60, 014012 (1999).
\bibitem{approach}  A.N. Sissakian, O.Yu. Shevchenko, A.P. Nagaytsev, O.N. Ivanov,  Phys. Rev. {\bf D72} (2005) 054027
 \bibitem{mul1} D. Boer, R. Jakob, P. J. Mulders,  Nucl. Phys. B {\bf  504}, 345 (1997)
 \bibitem{mul2} D. Boer, R. Jakob, P. J. Mulders, Phys. Lett. B {\bf 424}, 143 (1998) 
 \bibitem{mul3} D. Boer,  P. J. Mulders, Phys. Rev. D {\bf 57}, 5780 (1998) 


 \bibitem{efremov} A.V. Efremov et al, Phys. Lett. B {\bf 612}, 233 (2005).
 \bibitem{pythia} T. Sjostrand et al., hep-ph/0308153.  
\bibitem{conway} J.S. Conway et al, Phys. Rev. {\bf D39}, 92 (1989).
\bibitem{NA10} NA10 Collab., Z. Phys. {\bf C31}, 513 (1986);
 Z. Phys. {\bf C37}, 545 (1988).
 \bibitem{anselmino1} M. Anselmino, V. Barone, A. Drago, N.N. Nikolaev, Phys.Lett. {\bf B594}, 97 (2004).
\bibitem{bianconi} A. Bianconi, M. Radici Phys. Rev. {\bf D} 71, 074014 (2005).

\end{thebibliography}
\end{document}